\documentclass[aps,pra,twocolumn,floats]{revtex4-2} 
\usepackage{amsmath}
\usepackage{graphicx}
\usepackage{dcolumn}
\usepackage{bm}
\usepackage{amssymb}
\usepackage{latexsym}
\usepackage{color}
\usepackage{subfigure}
\usepackage{ulem}

\begin{document}

\title{Geometric Phases in Optics: Polarization of Light Propagating in Helical Optical Fibers}

\author{Y. B. Band$^{1,2,3}$, Igor Kuzmenko$^{1,2}$, Yshai Avishai$^{1,4}$}

\affiliation{
  $^1$Department of Physics,
  Ben-Gurion University of the Negev,
  Beer-Sheva 84105, Israel
  \\
  $^2$Department of Chemistry,
  Ben-Gurion University of the Negev,
  Beer-Sheva 84105, Israel
  \\
  $^3$The Ilse Katz Center for Nano-Science,
  Ben-Gurion University of the Negev,
  Beer-Sheva 84105, Israel
  \\
  $^4$Yukawa Institute of Theoretical Physics, Kyoto, Japan
  }

\begin{abstract}
The geometric phase in optics (GPIO) is directly associated with the polarization of light. We investigate the physical principles underlying the occurrence of the GPIO for a single-mode light beam propagating in a single-mode optical fiber wound into a circular helix configuration, with and without stress-induced birefringence.  The effects of the curvature and torsion of the helical fiber on the rotation of the polarization vector and the associated GPIO are discussed. Analytic expressions are derived for the polarization vector and Stokes parameters for any initial polarization state of the light entering the helical fiber, as well as for the GPIO of the light as a function of helix arc-length. Additionally, the intensity of a superposition of the initial and final beams, which depends on the final GPIO, is derived.  Furthermore, the relationship between the GPIO and the solid angle subtended by the tangent vector of the helix plotted on the Poincar\'{e} sphere is analyzed, and the effects of fluctuations of the parameters specifying the geometry and the material characteristics of the helical fiber on the GPIO are considered.
\end{abstract}

\maketitle

\section{Introduction} \label{sec:Intro}

The propagation of light in optical fibers is an important topic in optics. When an optical fiber is wound into a twisted configuration, such as a helix, the polarization state of the light propagating through the fiber can undergo a variety of interesting effects due to the geometric properties of the twisted fiber. One such effect arises from the optical geometric phase (GPIO), sometimes called the Pancharatnam-Berry geometric phase, $\gamma_{\rm PB}$, which emerges due to the non-trivial geometry of the fiber, as has been described in Refs.~\cite{Pancharatnam_56, Berry_87, Berry_87_N, Sorce_23, Berry_84, Simon_83, Ross_84, Wang_23, Mieling_23, Jisha_21, Wang_23, Hasman_03, Bomzon_02, Hasman_03, Marrucci_06, Tomita_86} and references therein.  This phase is acquired by a light beam when its polarization state undergoes cyclic evolution, i.e., the GPIO arises when a polarization vector evolves along a closed path in parameter space. For a circular helix optical fiber with and without significant stress-induced birefringence, the curvature and torsion of the coiled fiber play a crucial role in determining the GPIO.  As a consequence of the helical configuration, the polarization vector undergoes rotation as the light propagates. The mathematical formalism for calculating the rotation and the GPIO is related to the Berry connection; the GPIO is the integration of the Berry connection over the closed path in parameter space.  The GPIO can be experimentally quantified by splitting off a portion of the light entering and exiting the fiber and forming a superposition of these portions of the beam, and then measuring the resulting light intensity.  

Phase has broad relevance in optics; for example, Ref.~\cite{Sorce_23} deals with the formation of geometric phases during non-adiabatic frequency-swept radio frequency (RF) pulses with sine amplitude modulation and cosine frequency modulation functions. The study analyzes the geometric phases using a Schr\"odinger equation formalism and provides solutions for sub-geometric phase components. The results have implications for magnetic resonance imaging (MRI) and high-resolution nuclear magnetic resonance (NMR).  Here we study the physical principles underlying the propagation of polarized light and the GPIO in helical optical fibers.

A considerable amount of research has been devoted to the GPIO, see Refs.~\cite{Berry_87, Berry_87_N, Berry_84, Simon_83, Wang_23, Mieling_23, Bomzon_02, Hasman_03, Marrucci_06, Tomita_86} and references therein, with applications in various optical systems. Reference~\cite{Berry_84} develops the theoretical framework of geometric phases with particular emphasis on what is now often called the Berry phase, which arises during adiabatic changes in quantum systems. Geometric phases depend on the path traversed in parameter space and have important implications in quantum mechanics \cite{Wu-Yang-75, Band-Avishai-QuantumMechanics}, mechanics in general, and optics.  Reference~\cite{Simon_83} emphasizes its connection of holonomy (the failure of parallel transport around closed cycles to preserve the geometric information being transported), the Berry phase and the quantum adiabatic theorem; it provides mathematical insights into the geometric phase and its importance in quantum physics and beyond.

The rotation of the polarization vector of a light beam propagating in an optical fiber has practical applications, including \cite{Jisha_21}:
\begin{enumerate}
\item Optical sensing: The GPIO can be used for sensing applications. Changes in the curvature or torsion of the fiber can be detected by monitoring the polarization state \cite{Wang_23}.
\item Optical communications: Helical fibers can be used to create polarization-based devices such as polarization rotators or mode converters \cite{Mieling_23}.
\item A spin-dependent deflector based on the GPIO has been realized for IR wavelengths in metallic nano-gratings working in transmission \cite{Bomzon_02}.
\item Lenses based on the GPIO \cite{Hasman_03}.
\item The intrinsic spin-orbit interaction associated with a gradient in the GPIO can produce a novel type of electromagnetic angular momentum transfer \cite{Bomzon_02, Marrucci_06}.
\item Quantum optics: The GPIO is relevant in quantum optics, where geometric phases play a crucial role \cite{Bomzon_02, Marrucci_06}.
\end{enumerate}

The GPIO in twisted optical fibers provides a rich playground for exploring geometric effects in optics. Understanding this phase can lead to novel applications and improve our ability to manipulate light polarization in optical systems.

The following papers are also particularly relevant to the work reported here.  
Reference \cite{Ross_84} investigates the modification of polarization in low and high stress-induced birefringence mono-mode optical fibers due to geometric effects. Specifically, it examines how bending and curving the fiber path can lead to polarization rotation.
Reference \cite{Mieling_23} analyzes the propagation of light in arbitrarily curved step-index optical fibers. Using a multi-scale approximation scheme, they rigorously derive Rytov's law (the rotation of the plane of polarization of an electromagnetic wave in a uniform inhomogeneous medium due to the GPIO) at leading order, and explore nontrivial dynamics of the polarization of the electromagnetic field. It describes how the curvature and torsion of the optical fiber affect the polarization vector of the light (this is an inverse Spin Hall effect for light \cite{Mieling_23}).
Measurement of the GPIO have been reported in Refs.~\cite{Ross_84, Tomita_86, Hannonen_20, Ferrer-Garcia_23}.

The outline of this manuscript is as follows: Section~\ref{sec:geometric_phase} presents a brief introduction to geometric phase in mechanics, quantum mechanics and optics.  Section~\ref{sec:geometry_helix} discusses the geometry of a circular helical optical fiber and the Frenet-Serret system (the basis vectors and the generalized curvatures for a curve in $\mathcal{R}^n$) [in $\mathcal{R}^3$ the basis vectors are the tangent vector {\bf T}, the normal ${\bf N}$, and the binormal vector ${\bf B}$, which form an orthonormal basis, and the curvature $\kappa$ and the torsion $\tau$ parameters].  Section~\ref{sec:polarization_helix} considers the dynamics of light polarization in a helical fiber as the light propagates along the fiber, and presents the equations of motion for amplitudes of the normal and binormal vectors in the light polarization vector that propagates in the helical fiber, Sec.~\ref{subsec:Stokes_Poincare} plots the Stokes vector on the Poincar\'{e} sphere, and Sec.~\ref{subsec:no_birefringence} considers the dynamics without any stress-induced birefringence.  
Sec.~\ref{sec:GPIO} introduces the GPIO for a helical fiber, Secs.~\ref{subsec:dynamic_phase}, \ref{subsec:geometric_phase} present analytic formulas for the Pancharatnam-Berry phase (PBP) and the Frenet-Serret geometric phase (FSGP) and the total geometric phase, respectively, Sec.~\ref{subsec:Adiabatic_limit} considers the adiabatic limit of the geometric phases, and Sec.~\ref{subsec:dyn_geo_total_phase} presents the geometric phases versus initial conditions of the polarization vector.  Section~\Ref{sec:coherent_superposition} discusses the intensity of a coherent superposition of the initial and final polarization vectors.  Section~\ref{sec:gammaB-Omega} considers the relationship between $\gamma_{\rm FS}$ and $\Omega$. Section~\ref{sec:fluctuations} calculates the effects of fluctuations of the parameters in the expressions for the GPIO, and finally, Sec.~\ref{sec:summary} contains a summary and conclusion.  Appendix \ref{append:phases} addresses the representation of the PBP and FSGP in a helical fiber using eigenvectors of the matrix in the dynamical equation for the amplitudes of the normal and binormal vectors in the light polarization vector, and Appendix  \ref{append-gamma_dyn-max-min} considers the extrema of the PBP.  

\section{Geometric phase in mechanics, quantum mechanics and optics}  \label{sec:geometric_phase}

In the context of classical and quantum mechanics, the term ``geometric phase'' refers to a phase acquired by a system in the course of carrying out motion along a closed curve.  The geometric phase is frequently studied in systems that vary adiabatically, i.e. slowly, and it is derived by considering the geometric properties of the parameter space of a Hamiltonian system~\cite{Solem-geometric-phase-1993}.  In optics it is known as the optical geometric phase, but it is also sometimes called the Pancharatnam phase, the Berry phase, or the Pancharatnam-Berry phase. The phenomenon was independently discovered in classical optics by S. Pancharatnam in 1956 \cite{Pancharatnam_56} and by H. C. Longuet-Higgins in molecular physics in 1958 \cite{LH_58}.  A noteworthy manifestation can be observed in the Aharonov--Bohm effect \cite{AB_59}.  Two further examples of a geometric phase arise in the  Foucault pendulum, as explained in Ref.~\cite{WS_89}, and in nuclear magnetic resonance, see Ref.~\cite{WS_89b}.  One may observe the effects of a geometric phase as follows: place your right arm at your side, then move it straight in front of you by rotating it 90 degrees.  Next rotate it another 90 degrees so that your arm is extended to your right, and finally rotate your arm another 90 degrees so that your arm is back at your side.  Notice that your hand is rotated 90 degrees from its original direction.  This rotation of your hand is a geometric phase.

The polarization vector of the light is an important quantity in the field of optics.  We shall denote the slowly varying envelope of the electric field of the light by the symbol ${\bf A}$.  In what follows we shall consider the propagation of ${\bf A}$ as a function of the arc-length $s$ along the fiber, ${\bf A}(s)$.  The light propagates in the direction of the wave vector of the light, ${\bf k}$, which is guided along the fiber and therefore propagates along the tangent vector of the fiber, ${\bf T}(s)$.  The vector ${\bf A}(s)$ is normal to the vector ${\bf T}(s)$ and has components along the normal and binormal vectors, ${\bf N}(s)$ and ${\bf B}(s)$, i.e., ${\bf A}(s) = A_n(s) {\bf N}(s) + A_b(s) {\bf B}(s)$, where $A_n(s)$ and $A_b(s)$ are the amplitudes of the normal and binormal vectors.  In quantum mechanics \cite{Band-Avishai-QuantumMechanics}, the geometric phase is associated with amplitudes of basis functions which vary with time.  But in optics, not only do $A_n(s)$ and $A_b(s)$ vary with arc-length, but the basis vectors ${\bf N}(s)$ and ${\bf B}(s)$ also vary with arc-length.  As a consequence of this variation of the basis vectors with arc-length, there is an additional geometric phase that is not present in quantum mechanical treatments.  It is therefore necessary to discuss what can be thought of as a `dynamic geometric phase', the ``Pancharatnam-Berry phase (PBP)'', for which we introduce the symbol $\gamma_{\rm PB}(s)$, as well as {\it a new geometric phase} that arises from the variation of the basis vectors ${\bf N}(s)$ and ${\bf B}(s)$ with arc-length.  We call the latter phase the ``Frenet-Serret geometric phase (FSGP)'', for which we introduce the symbol $\gamma_{\rm FS}(s)$.  To the best of our knowledge, the FSGP has not been previously discussed in the literature.  We shall call the sum of the PBP and the FSGP, the {\it total geometric phase}.

\section{Geometry of an optical helix fiber}  \label{sec:geometry_helix}

A circular helix of radius $r$ and pitch $2 \pi c$ is defined as a vector-valued function in 3D space,
\begin{equation}  \label{eq:helix-z-vs-t}
  {\bf R} (t) =
  r \, \cos t \, \hat{\bf x} +
  r \, \sin t \, \hat{\bf y} +
  c \, t \, \hat{\bf z} ,
\end{equation}
where $\hat{\bf x}$, $\hat{\bf y}$ and $\hat{\bf z}$ are Cartesian basis vectors, the origin of the coordinate system is on the axis of the helix, $r$ and $c$ are assumed to be positive, and the real dimensionless curve parameter $t$ (you can think of the parameter $t$ as being the product of the beam phase velocity and the propagation time expressed in dimensionless units).  The arc-length $s$ of the helix is
\begin{equation}   \label{eq:arclength-helix}
  s = \sqrt{r^2 + c^2} \, t .
\end{equation} 
Solving Eq.~(\ref{eq:arclength-helix}) for $t$, we obtain $t = s / \sqrt{r^2 + c^2}$, i.e., we invert $s(t)$ to get $t(s)$.
Hence, the curve of the helix in Eq.~(\ref{eq:helix-z-vs-t}) can be re-parameterized to give ${\bf R}$ as a function of $s$:
\begin{eqnarray}
  {\bf R} (s) &=&
  r \cos \bigg( \frac{s}{\sqrt{r^2 + c^2}} \bigg) \, \hat{\bf x} +
  r \sin \bigg( \frac{s}{\sqrt{r^2 + c^2}} \bigg) \, \hat{\bf y}
  \nonumber \\ && +
  \frac{c \, s}{\sqrt{r^2 + c^2}} \, \hat{\bf z} .
  \label{eq:helix-z}
\end{eqnarray}
Note that $| d {\bf R} (s) / ds | = 1$, hence ${\bf T}(s) = d {\bf R}(s)/ds$ is the tangent vector of unit length.  In the helical case,
\begin{eqnarray}
  {\bf T}(s) &=&
  \frac{1}{\sqrt{r^{2} + c^2}} \,
  \bigg[
    - r \sin \bigg( \frac{s}{\sqrt{r^2 + c^2}} \bigg) \, \hat{\bf x}
    \nonumber \\ && +
    r \cos \bigg( \frac{s}{\sqrt{r^2 + c^2}} \bigg) \, \hat{\bf y} +
    c \, \hat{\bf z}
  \bigg] .
  \label{eq:T-def}
\end{eqnarray}
The unit normal vector ${\bf N} (s)$ can be obtained from the equation $d {\bf T} (s) / d s = \kappa \, {\bf N} (s)$, where the curvature $\kappa$ is given by
\begin{equation}   \label{eq:curvature}
  \kappa = \Big| \frac{d {\bf T}(s)}{d s} \Big| = \frac{r}{r^{2} + c^2},
\end{equation}
and using these results we find:
\begin{equation}  \label{eq:normal-vector}
  {\bf N} (s) =
  - \cos \bigg( \frac{s}{\sqrt{r^2 + c^2}} \bigg) \, \hat{\bf x}
  - \sin \bigg( \frac{s}{\sqrt{r^2 + c^2}} \bigg) \, \hat{\bf y} .
\end{equation}
The binormal vector is given by ${\bf B} (s) = {\bf T} (s) \times {\bf N} (s)$:
\begin{eqnarray}
  {\bf B} (s) &=&
  \frac{1}{\sqrt{r^2 + c^2}} \,
  \bigg[
    c \sin \bigg( \frac{s}{\sqrt{r^2 + c^2}} \bigg) \, \hat{\bf x}
    \nonumber \\ && -
    c \cos \bigg( \frac{s}{\sqrt{r^2 + c^2}} \bigg) \, \hat{\bf y} +
    r \, \hat{\bf z}
  \bigg] .
  \label{eq:bi-normal-vector}
\end{eqnarray}
By differentiating ${\bf B} (s)$ we find $d {\bf B} (s) / d s = - \tau \, {\bf N} (s)$, where the torsion of the helix is $\tau = - {\bf N} (s) \cdot d {\bf B} / d s$:
\begin{equation}   \label{eq:torsion}
  \tau = \frac{c}{r^{2} + c^2} .
\end{equation}

More generally, the Frenet--Serret formulas \cite{FSS} describe differentiable curves in ${\mathcal{R}}^n$.  Although the curve ${\bf R}(t)$ is given as a function of the dimensionless curve parameter $t$, the  Frenet--Serret system is usually written in terms of the arc-length $s$:
\begin{equation}
  {\bf R} (s) = x_1(s) \, \hat{\bf x}_1 + x_2(s) \, \hat{\bf x}_2 + \ldots + x_n(s) \, \hat{\bf x}_n .
\end{equation}
Here $x_i(s)$ are smooth functions of the arc-length $s$, and the unit tangent vector is given by ${\bf T} (s) = \frac{d {\bf R} (s)}{d s}$.  In ${\mathcal{R}}^n$ there are $n-1$ generalized curvatures, and $n$ Frenet--Serret unit basis vectors.

As mentioned above, in three-dimensional Euclidean space, ${\mathcal{R}}^3$, the normal vector ${\bf N} (s)$ and the binormal vector ${\bf B} (s)$ are given by
\begin{eqnarray}
  {\bf N} (s) &=&
  \frac{1}{\kappa (s)} \, \frac{d {\bf T} (s)}{d s} ,
  \label{eq:N-general}
  \\
  {\bf B} (s) &=&
  {\bf T} (s) \times {\bf N} (s) ,
  \label{eq:B-general}
\end{eqnarray}
The curvature $\kappa (s)$ measures the failure of the curve ${\bf R} (s)$ to be a straight line, while torsion $\tau (s)$ measures the failure of the curve to be planar:
\begin{eqnarray}
  \kappa (s) &=&
  \frac{| {\bf R}' (s) \times {\bf R}'' (s)|}{| {\bf R} (s)|^{3}} ,
  \label{eq:curvature-general}
  \\
  \tau (s) &=&
  \frac{[{\bf R}' (s) \times {\bf R}'' (s)] \times {\bf R}''' (s)}{|{\bf R}' (s) \times {\bf R}'' (s)|^{2}} ,
  \label{eq:torsion-general}
\end{eqnarray}
where ${\bf R}' (s) = d {\bf R} (s) / d s$, ${\bf R}'' (s) = d^2 {\bf R} (s) / d s^2$ and ${\bf R}''' (s) = d^3 {\bf R} (s) / d s^3$.  The Frenet--Serret system describes the curvature $\kappa (s)$ and torsion $\tau (s)$ of the curve and the derivatives of the tangent, normal, and binormal unit vectors in terms of each other,
\begin{equation}   \label{eq:Frenet--Serret-system}
\begin{aligned}
\frac{d {\bf T}}{ds}&=\kappa {\bf N} ,\\
\frac{d {\bf N}}{ds}&=-\kappa {\bf T} +\tau {\bf B} ,\\
\frac{d {\bf B}}{ds}&=-\tau {\bf N} .
\end{aligned}
\end{equation}
The unit vectors ${\bf T}$, ${\bf N}$, ${\bf B}$, and the scalars $\kappa$ and $\tau$, are together called the Frenet--Serret apparatus.  [In the {\it Wolfram Mathematica} language used to carry out the calculations reported here, the Frenet--Serret system is called `FrenetSerretSystem'; the Frenet--Serret equations are coded as a function of a parameter $t$, which is not necessarily the arc-length $s$.]  We denote the period of the tangent, normal and binormal vectors by $P_{\rm FS}$:
\begin{equation}   \label{eq:period_FS}
P_{\rm FS} = 2 \pi \sqrt{r^2 + c^2} = \frac{2 \pi}{\sqrt{\kappa^2 + \tau^2}}.
\end{equation}

\begin{figure}
\centering
  \includegraphics[width = 0.9 \linewidth,angle=0] {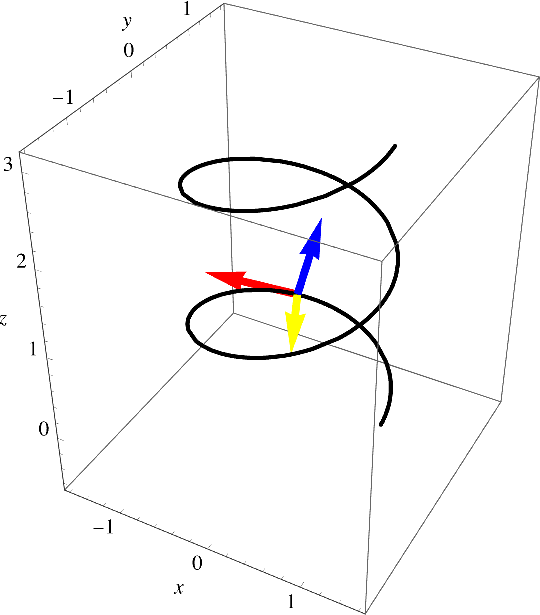}
\caption{\footnotesize  
Helix ${\bf R}(s)$ (black), and ${\bf T}(s)$ (red), ${\bf N}(s)$ (yellow), ${\bf B}(s)$ (blue) plotted with parameters $r = 1$ and $c = 1/4$. The triad of unit vectors are plotted for $s = 2$.
}
\label{Fig:helix_T_N_B}
\end{figure}

\begin{figure}
\centering
  \includegraphics[width = 0.9 \linewidth,angle=0] {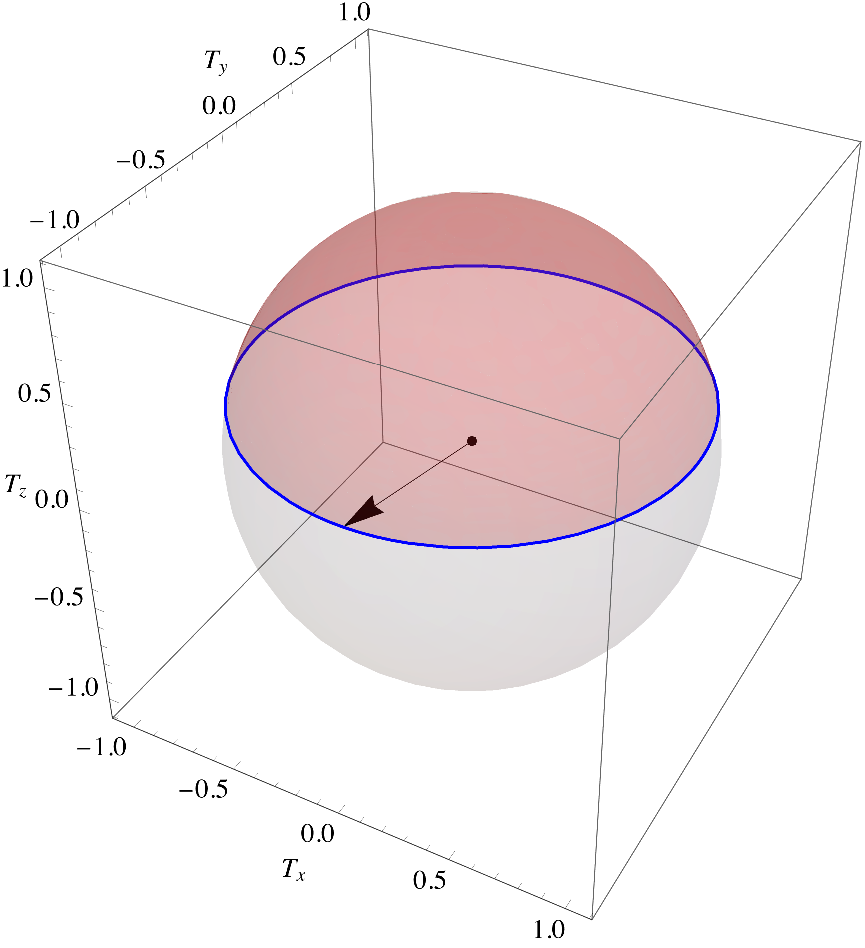}
\caption{\footnotesize  
Tangent vector ${\bf T}(s)$ on the Poincar\'{e} sphere plotted with parameters $r = 1$ and $c = 1/4$.    The arrow shows the initial value, ${\bf T}(0)$.  The solid angle $\Omega(r,c)$ is depicted by semi-transparent red.}
\label{Fig:T_on_Poincare_sphere}
\end{figure}

Figure \ref{Fig:helix_T_N_B} shows a helix ${\bf R}(s)$ with parameters $r = 1$ and $c = 1/4$ plotted in black, and the unit vectors ${\bf T}(s)$ (red), ${\bf N}(s)$ (yellow), ${\bf B}(s)$ (blue) evaluated at $t = 2$ ($s = 2\sqrt{17/16}$.  A triad of unit vectors can be plotted at any value of arc-length $s$.

Figure \ref{Fig:T_on_Poincare_sphere} shows the tangent vector ${\bf T}(s)$ on the unit sphere (i.e., the Poincar\'{e} sphere) as a blue curve, where the unit vector ${\bf T}(s)$ has its tail at the origin and its head on the blue curve on the Poincar\'{e} sphere.  The figure is drawn for helix parameters $r = 1$ and $c = 1/4$.  The solid angle on the sphere subtended by ${\bf T}(s)$ as $s$ moves from $s = 0$ to $s = P_{\rm FS}$ is drawn as semi-transparent red.  We call this solid angle $\Omega(r,c)$:
\begin{equation}   \label{eq:Omega}
  \Omega (r, c) = \frac{2 \pi \, c}{\sqrt{r^2 + c^2}} = \frac{2 \pi \, \tau}{\sqrt{\kappa^2 + \tau^2}}.
\end{equation}
Note that as $r$ decreases (increases) the area decreases (increases), and as $c$ decreases (increases), the solid angle $\Omega(r,c)$ increases (decreases).

\section{Dynamics of light polarization in a helix fiber}  \label{sec:polarization_helix}

For a fiber wound into a circular helix (e.g., by wrapping the fiber around a cylinder), an analytic solution for the polarization state of the light as a function of arc-length in a fiber with both significant and negligible stress-induced birefringence is possible as both $r$ and $c$ (and therefore $\kappa$ and $\tau$) are constants independent of $s$.  The propagation equations for the polarization vector
\begin{equation}  \label{eq:A_vector}
  {\bf A}(s) = A_n(s) {\bf N}(s) + A_b(s) {\bf B}(s), 
\end{equation}
where the subscripts $n$ and $b$ stand for normal and binormal, is given by \cite{Ross_84, Jones_48}
\begin{equation}  \label{eq:amp_eq}
i \, \frac{d}{ds} \left( \! \!
\begin{array}{c}
  A_n(s) \\
  A_b(s)
\end{array}  \! \!\right) = 
\left(  \! \!
\begin{array}{cc}
  -\frac{\alpha \, \kappa^2}{2} & i \tau    \\
  - i \tau &  \frac{\alpha \, \kappa^2}{2}   
\end{array}
 \! \! \right) \left(  \! \!
\begin{array}{c}
  A_n(s)  \\
  A_b(s)
\end{array}  \! \! \right) .
\end{equation}
Here the parameter $\alpha$ is the product of the square of radius of the optical fiber, $\rho$, and a constant $a$ depending on the wavelength of the light and the elastic material properties of the fiber that give rise to stress-induced birefringence, $\alpha = \rho^2 a$  \cite{Ross_84, Jones_48}.  The solution to Eq.~(\ref{eq:amp_eq}) with the general initial conditions
\begin{equation}    \label{eq:initial}
  A_n(0) = p , \qquad  A_b(0) = e^{i \phi} \sqrt{1 - p^2} ,
\end{equation}
where $p$ is the probability amplitude for being along the normal vector and $\phi$ is the initial phase of the amplitude $A_b(0)$, takes the form
\begin{eqnarray}
  A_n(s) &=&
  p \, \cos (k_s s) +
  \frac{2 \tau e^{i \phi} \sqrt{1 - p^2} + i \, \alpha \kappa^2 p }{2 k_s} \,
  \sin (k_s s) ,
   \nonumber \\ 
  A_b(s) &=&
  e^{i \phi} \sqrt{1 - p^2} \cos (k_s s)
  \nonumber \\ &-&
  \frac{2 \tau p + i \, \alpha \kappa^2 e^{i \phi} \sqrt{1 - p^2}}{2 k_s} \,
  \sin (k_s s) ,
\label{eq:An-Ab-vs-s}
\end{eqnarray}
where $k_s = \frac{1}{2} \sqrt{\alpha^2 \kappa^4 + 4 \tau^2}$.

\begin{figure}
\centering
  \includegraphics[width = 0.9 \linewidth,angle=0] {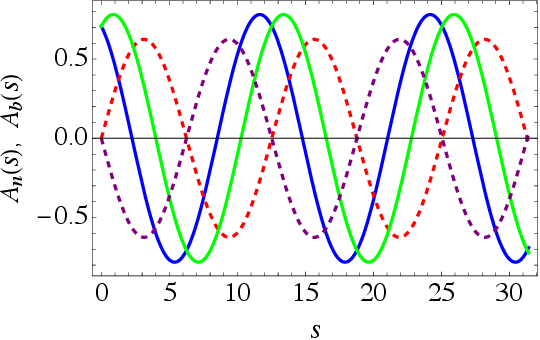}
\caption{\footnotesize  
The amplitudes $A_n(s)$ and $A_b(s)$ of the normal and binormal vectors in the polarization vector ${\bf A}(s)$ [see Eq.~(\ref{eq:A_vector})] for $p = 1/\sqrt{2}$ and $\phi = 0$ (linearly polarized light with polarization at a 45$^{\circ}$ angle from the $x$-axis).  Re$[A_n(s)]$ is the solid blue curve, Im$[A_n(s)]$ is the dashed red curve, Re$[A_b(s)]$ is the solid green curve, and Im$[A_b(s)]$ is the dashed purple curve.
}
\label{Fig:A_n-A_b.P=0.5.phi=0}
\end{figure}

\begin{figure}
\centering
  \includegraphics[width = 0.9 \linewidth,angle=0] {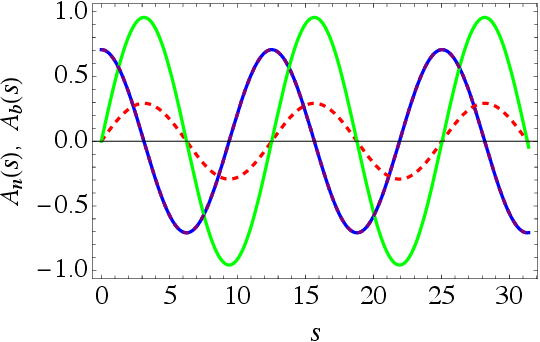}
\caption{\footnotesize  
The amplitudes $A_n(s)$ and $A_b(s)$ of the normal and binormal vectors in the polarization vector ${\bf A}(s)$ [see Eq.~(\ref{eq:A_vector})] for left circular polarized light, $p = 1/\sqrt{2}$ and $\phi = \pi/2$.  The Re$[A_n(s)]$ is the solid blue curve, Im$[A_n(s)]$ is the dashed red curve, Re$[A_b(s)]$ is the solid green curve, and Im$[A_b(s)]$ is the dashed purple curve which overlaps the solid blue Re$[A_n(s)]$ curve.
}
\label{Fig:A_n-A_b.P=0.5.phi=pi/2}
\end{figure}

\begin{figure}
\centering
  \includegraphics[width = 0.9 \linewidth,angle=0] {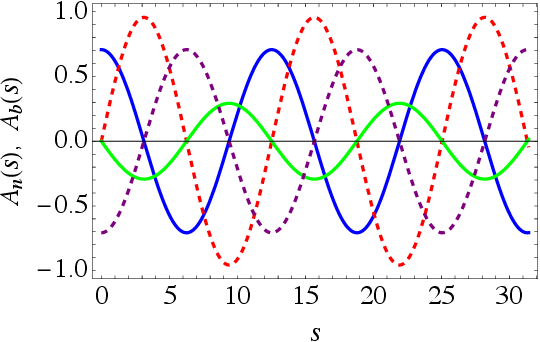}
\caption{\footnotesize  
Amplitudes  $A_n(s)$ and $A_b(s)$ of the normal and binormal vectors in the polarization vector ${\bf A}(s)$ for right circular polarized light, $p = 1/\sqrt{2}$ and $\phi = -\pi/2$.  The Re$[A_n(s)]$ is the solid blue curve, Im$[A_n(s)]$ is the dashed red curve, Re$[A_b(s)]$ is the solid green curve, and Im$[A_b(s)]$ is the dashed purple curve.
}
\label{Fig:A_n-A_b.P=0.5.phi=-pi/2}
\end{figure}

Figure \ref{Fig:A_n-A_b.P=0.5.phi=0} plots the amplitudes $A_n(s)$ and $A_b(s)$ versus $s$ for linearly polarized light with polarization at a 45$^{\circ}$ angle from the $x$-axis, i.e., probability amplitude $p = 1/\sqrt{2}$ and the phase $\phi = 0$.    The parameter values used are $\alpha = 1$, $r = 1$ and $c = 1/4$.  The amplitudes have a period $P_s = 2 \pi/k_s =  4 \pi/\sqrt{\alpha ^2 \kappa ^4 + 4 \tau ^2}$. The curves Im$[A_n(s)]$ and Im$[A_b(s)]$ are 180$^{\circ}$ out of phase; Re$[A_n(s)]$ and Re$[A_b(s)]$ are out of phase by a much smaller angle.

Figure \ref{Fig:A_n-A_b.P=0.5.phi=pi/2} plots the amplitudes $A_n(s)$ and $A_b(s)$ versus $s$ for left circular polarized light, i.e., probability amplitude $p = 1/\sqrt{2}$ and the phase $\phi = \pi/2$.  In this case, Re$[A_n(s)] =$ Im$[A_b(s)]$ so these curves lie one on top of the other.  Figure \ref{Fig:A_n-A_b.P=0.5.phi=-pi/2} is somewhat similar to Fig.~\ref{Fig:A_n-A_b.P=0.5.phi=pi/2}, except for right circular polarized light, $p = 1/\sqrt{2}$ and $\phi = -\pi/2$; here the Re$[A_n(s)]$ and Im$[A_b(s)]$ curves are distinct.

\subsection{Polarization vector and Stokes vector on the Poincar\'{e} sphere}    \label{subsec:Stokes_Poincare} 

In order to demonstrate that the dynamics of the polarization vector in a helical optical fiber can be represented by the polarization evolution on the Poincar\'{e} sphere, it is convenient to rewrite Eq.~(\ref{eq:amp_eq})  in a manner that resembles a dynamical equation for a spin-1/2 particle in a constant magnetic field:
\begin{equation}   \label{eq:amp_eq-Pauli}
  i \, \frac{d}{d s}
  \left(\! \!
    \begin{array}{c}
      A_n(s) \\
      A_b(s)
    \end{array}
  \! \!\right) = 
  \mathbb{B} \cdot \boldsymbol\sigma \,
  \left(\! \!
    \begin{array}{c}
      A_n(s)  \\
      A_b(s)
    \end{array}
  \! \! \right) .
\end{equation}
Here $\boldsymbol\sigma = (\sigma_z, \sigma_x, \sigma_y)$ [note the order], and the $s$-independent {\it fictitious} magnetic field vector $\mathbb{B}$ is given by
\begin{equation}  \label{eq:mathcalB}
  \mathbb{B} = \Big( - \frac{1}{2} \, \alpha \kappa^2 , \, 0 , \, - \tau \Big) .
\end{equation}
This notation is employed to elucidate  the analogy between the dynamics of a polarization vector in a helical optical fiber and the dynamics of a spin-1/2 particle in the presence of a {\it real} magnetic field.

For understanding the physics of the polarization, it is convenient to define the Stokes vector ${\bf S}_{\rm FS}(s) = (S_1(s), S_2(s), S_3(s))$ where $S_1, S_2$ and $S_3$ are the Stokes parameters.  The third component is along the propagation direction of the light, $\hat {\bf k} = {\bf T}(s)$  \cite{Jackson_99}.   ${\bf S}_{\rm FS}(s)$ is a real vector that can be plotted on the Poincar\'{e} sphere (as can the complex polarization vector ${\bf A}(s)$ be plotted on the complex Bloch sphere \cite{Band_L&M}):
\begin{eqnarray}
  {\bf S}_{\rm FS}(s) &=&  \big(
    A_n^{*} (s), A_b^* (s)
  \big) \,
  \boldsymbol\sigma \,
  \left(\! \!
    \begin{array}{c}
      A_n(s)  \\
      A_b(s)
    \end{array}
  \! \! \right) .
\label{s_vec}
\end{eqnarray} 
Here the Pauli vector $\boldsymbol\sigma$ is defined as $\boldsymbol\sigma = (\sigma_z, \sigma_x, \sigma_y)$ [note the order], where $\sigma_x$, $\sigma_y$ and $\sigma_z$ are the usual Pauli matrices.   

\begin{figure}
\centering
  \includegraphics[width = 0.9 \linewidth,angle=0] {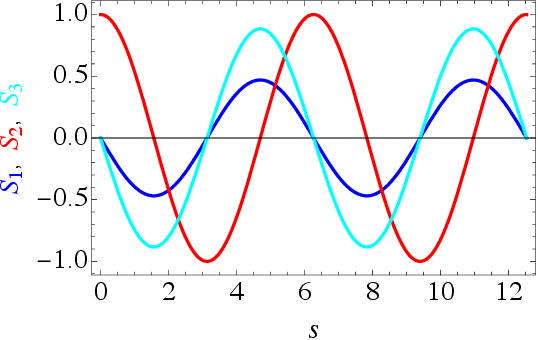}
\caption{\footnotesize  
The Stokes vector components of ${\bf S}_{\mathrm{FS}} (s)$ in the Frenet--Serret frame versus arc-length $s$ for initial conditions $p = 1/\sqrt{2}$ and $\phi = 0$ with with parameters $r = 1$ and $c = 1/4$.
\label{Fig:S_vs_t}
}
\end{figure}

Figure~\ref{Fig:S_vs_t} plots the components of ${\bf S}_{\mathrm{FS}} (s)$ versus $s$ for the initial linearly polarized case where the amplitudes $A_n(s)$ and $A_b(s)$  are shown in Fig.~\ref{Fig:A_n-A_b.P=0.5.phi=0}.  The period of Stokes vector  ${\bf S}_{\mathrm{FS}} (s)$ is half of the period $P_s$ of the amplitudes, where $P_s = 12.5281$, as can be seen in all the figures for the amplitudes shown above.

\begin{figure}
\centering
  \includegraphics[width = 0.9 \linewidth,angle=0] {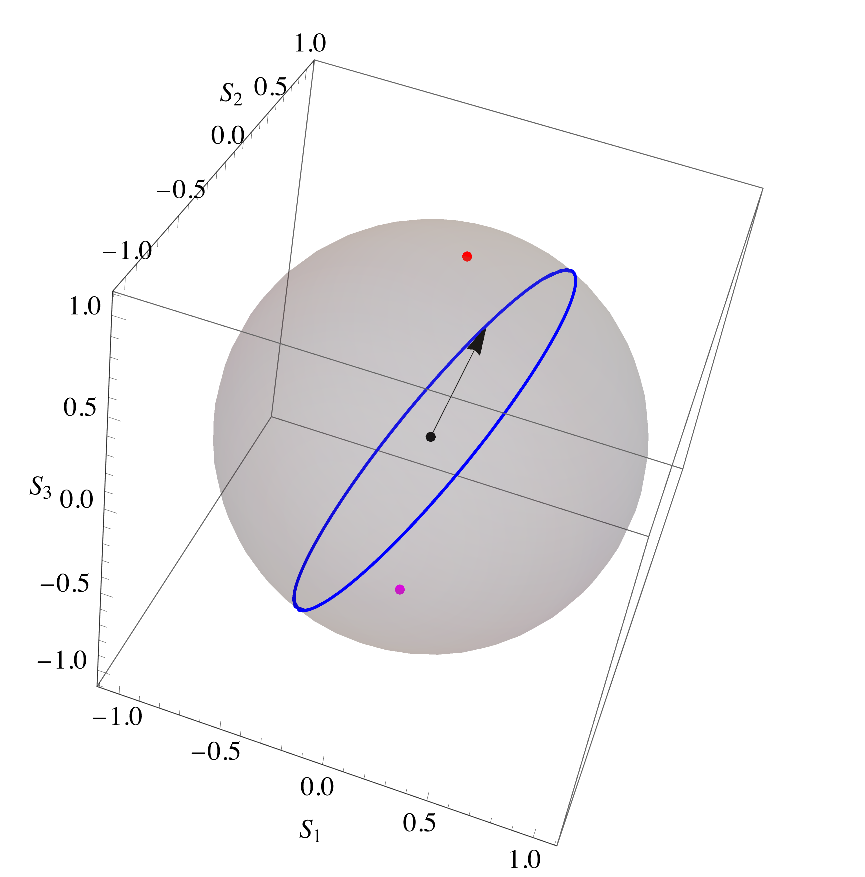}
\caption{\footnotesize  
The trajectory of the Stokes vector ${\bf S}_{\mathrm{FS}} (s)$ in the Frenet--Serret frame (blue curve) plotted on the Poincar\'{e} sphere for the initial conditions with $p = 1/\sqrt{2}$ and $\phi = 0$ (linearly polarized beam).  The arrow shows the initial value, ${\bf S}_{\mathrm{FS}} (0)$, the red point at the north pole indicates the instantaneous direction of ${\bf T}(s)$ and the magenta point at the south pole is the direction of $-{\bf T}(s)$.
\label{Fig:S_Poincare}
}
\end{figure}

Figure~\ref{Fig:S_Poincare} shows the dynamics of the Stokes vector ${\bf S}_{\mathrm{FS}} (s)$ in the Frenet--Serret frame on the Poincar\'{e} sphere for the linearly polarized case, $p = 1/\sqrt{2}$ and $\phi = 0$. This dynamics is analogous to the dynamics of a spin-1/2 particle on the Bloch sphere due to a magnetic field.  The black arrow shows the polarization vector at $s = 0$.  As propagation proceeds, the tip of  ${\bf S}_{\mathrm{FS}} (s)$ follows the red curve shown in the figure, making one full revolution at $s = P_s/2$ and two full revolutions at $s = P_s$.  As the initial conditions change the solid angle subtended by ${\bf S}_{\mathrm{FS}} (s)$ in the Frenet--Serret frame on the Poincar\'{e} sphere changes; as $p$ increases from zero to one, the solid angle subtended increases. When $p = 0$ and $p = 1$, there is no dependence on $\phi$, but when $p \ne 0$, first the solid angle increases with increasing $p$, then decreases, and for certain values of $\phi$ it can again increase.  The solid angle subtended equals the GPIO.  More will be said about the GPIO in Sec.~\ref{sec:GPIO}.

Note that the plot of the Stokes vector  in the Frenet--Serret frame ${\bf S}_{\mathrm{FS}} (s)$  on the Poincar\'{e} sphere is equivalent to the plot of the polarization vector in the Frenet--Serret frame on the Poincar\'{e} sphere.

\begin{figure}
\centering
  \includegraphics[width = 0.9 \linewidth,angle=0] {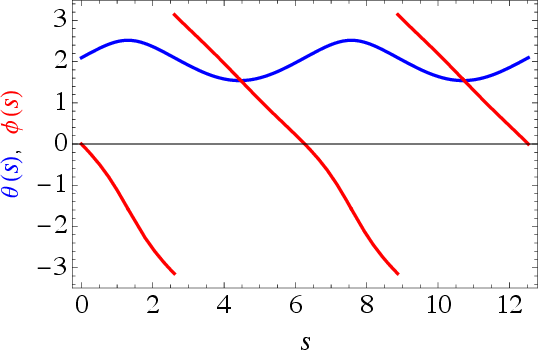}
\caption{\footnotesize  
The polar and azimuthal angles of the Stokes vector ${\bf S}_{\mathrm{FS}} (s)$, $\theta(s)$ and $\phi(s)$, versus $s$.
\label{theta_phi_vs_s}
}
\end{figure}

The position of the Stokes vector ${\bf S}_{\mathrm{FS}} (s)$ on the Poincar\'{e} sphere can be quantified by the polar angle $\theta(s)$ and azimuthal angle $\phi(s)$. Figure \ref{theta_phi_vs_s} plots these angles versus $s$ for the linear polarization case with $p = 1/\sqrt{2}$ and $\phi = 0$.  The discontinuities in $\phi(s)$ arises because of the way the azimuthal angle was calculated; it ranges from $-\pi$ to $\pi$.  Clearly these angles are also periodic.

It is also possible to define the Stokes vector in the space-fixed frame, ${\bf S} (s) = {\mathcal{R}}({\bf T}(s)) \, {\bf S}_{\rm FS} (s)$, where ${\mathcal{R}}({\bf T}(s))$ is the 3D rotation matrix that rotates the light propagation direction $(0, 0, 1)$ into the vector ${\bf T}(s)$.

\subsection{Dynamics without stress-induced birefringence}    \label{subsec:no_birefringence} 

Reference \cite{Berry_87_N} considers the dynamics of light propagation in the absence of stress-induced birefringence.  In this case, the parameter  $\alpha$ appearing in Eq.~(\ref{eq:amp_eq}) takes the value $\alpha = - \tfrac{\lambda}{4\pi}$, where $\lambda = \lambda_0/n$, $\lambda_0$ is the vacuum wavelength and $n$ is the fiber refractive index. The matrix appearing on the right-hand-side of Eq.~(\ref{eq:amp_eq}) then becomes
\begin{equation*}
\left(  \! \!
\begin{array}{cc}
  -\frac{\alpha \, \kappa^2}{2} & i \tau    \\
  - i \tau &  \frac{\alpha \, \kappa^2}{2}   
\end{array}
 \! \! \right) \to \left(  \! \!
\begin{array}{cc}
  \tfrac{\lambda \kappa^2}{8\pi} & i \tau    \\
  - i \tau &  -\tfrac{\lambda \kappa^2}{8\pi}   
\end{array}
 \! \! \right).
\end{equation*}
Clearly, the analytic solution in Eq.~(\ref{eq:An-Ab-vs-s}) remains valid when the stress-induced birefringence vanishes so $\alpha = - \tfrac{\lambda}{4\pi}$.  [Reference \cite{Berry_87_N} adds a 2$\times$2 unit matrix $\tfrac{\lambda \kappa^2}{8\pi} {\bf 1}$ to this matrix so the (2,2) element of the matrix vanishes, and then the analytic solution in  Eq.~(\ref{eq:An-Ab-vs-s}) is multiplied by an additional phase factor $e^{- i \lambda \kappa^2 s/(8\pi)}$].

\section{PBP and FSGP in a helical fiber}  \label{sec:GPIO}

The total geometric phase $\gamma (s)$ (i.e., the sum of the PBP and FSGP) as a function of the arc-length $s$
can be found from the equation
\begin{equation}   \label{eq:gamma-vs-A}
  \frac{d \gamma (s)}{d s} = i {\bf A}^{*} (s) \cdot \frac{d {\bf A} (s)}{d s} .
\end{equation}
Using Eq.~(\ref{eq:Frenet--Serret-system}), $d \gamma (s) / d s$ can be written as
\begin{equation}   \label{eq:gamma=dyn+geo}
  \frac{d \gamma (s)}{d s} = \frac{d \gamma_{\rm PB} (s)}{d s} + \frac{d \gamma_{\rm FS} (s)}{d s} ,
\end{equation}
where
\begin{eqnarray}
  \frac{d \gamma_{\rm PB} (s)}{d s} &=&
  i \, A_{n}^{*} (s) \, \frac{d A_n (s)}{d s} +
  i \, A_{b}^{*} (s) \, \frac{d A_b (s)}{d s}
  \nonumber \\ &=&
  \big( A_{n}^{*} (s) , A_{b}^{*} (s) \big) \,
  \left(
    \begin{array}{cc}
      - \frac{\alpha \kappa^2}{2} & i \tau
      \\
      - i \tau & \frac{\alpha \kappa^2}{2}
    \end{array}
  \right) \,
  \left(
    \begin{array}{c}
      A_n (s) \\ A_b (s)
    \end{array}
  \right) ,
  \nonumber \\
  \label{eq:d-gamma-dyn}
  \\
  \frac{d \gamma_{\rm FS} (s)}{d s} &=&
  \big( A_{n}^{*} (s) , A_{b}^{*} (s) \big) \,
  \left(
    \begin{array}{cc}
      0 & - i \tau
      \\
      i \tau & 0
    \end{array}
  \right) \,
  \left(
    \begin{array}{c}
      A_n (s) \\ A_b (s)
    \end{array}
  \right) .  \label{eq:d-gamma-geo}
\end{eqnarray}
We shall refer to the first term on the right-hand side of Eq.~(\ref{eq:gamma=dyn+geo}), $d \gamma_{\rm PB} (s) / d s$, as a PBP (see the discussion in the Sec.~\ref{subsec:dynamic_phase} below), and to the second term, $d \gamma_{\rm FS} (s) / d s$, as the FSGP; the latter arises from the second term on the right-hand-side of Eq.~(\ref{eq:gamma=dyn+geo}) which can be written as $i {\bf A}^{*} (s) \cdot [A_n (s) \, \tfrac{d {\bf N} (s)}{ds} + A_b (s) \, \tfrac{d{\bf B} (s)}{ds}]$.

\subsection{Pancharatnam-Berry phase}  \label{subsec:dynamic_phase}

Using Eqs.~(\ref{s_vec}) and (\ref{eq:d-gamma-dyn}), we can express $d \gamma_{\rm PB} (s) / d s$ as
\begin{equation}   \label{eq:d-gamma_dyn-vs-S}
  \frac{d \gamma_{\rm PB} (s)}{d s} = \mathbb{B} \cdot {\bf S}_{\rm FS}(s) ,
\end{equation}
where $\mathbb{B}$ is defined in Eq.~(\ref{eq:mathcalB}) and the Stokes vector ${\bf S}_{\rm FS}(s)$ is defined in Eq.~(\ref{s_vec}).  Substituting Eq.~(\ref{eq:An-Ab-vs-s}) into Eq.~(\ref{s_vec}), one can see that $\mathbb{B} \cdot {\bf S}_{\rm FS}(s)$ does not depend on $s$, therefore $d \gamma_{\rm PB} (s) / d s$ does not depend on $s$, hence $\gamma_{\rm PB} (s)$ is a linear function of $s$:
\begin{equation}   \label{eq:gamma-dyn-Omega-p-phi}
  \gamma_{\rm PB} (s) = [(1 - 2 p^2) \alpha \kappa^2 - 4 p \sqrt{1 - p^2} \, \tau \sin \phi ] \, \frac{s}{2} .
\end{equation}

The extrema of $\gamma_{\rm PB} (s)$ (for a given value of $s$) over $p$ and $\phi$ are
(see Appendix~\ref{append-gamma_dyn-max-min} for details)
\begin{eqnarray}
  \gamma_{\rm PB,max} (s) &=& \frac{\sqrt{\alpha^2 \kappa^4 + 4 \tau^2}}{2} \, s ,
  \label{eq:dyn_max}
  \\
  \gamma_{\rm PB,min} (s) &=& -  \gamma_{\rm PB,max} (s) .
  \label{eq:dyn_min}
\end{eqnarray}
The two (equal) maxima of $\gamma_{\rm PB} (P_{\rm FS})$ occur for $p$ and $\phi$ given by
\begin{eqnarray*}
  ( p, \phi )
  &=&
  \pm
  \Big(
    \sqrt{\frac{\sqrt{\alpha^2 \kappa^4 + 4 \tau^2} - \alpha \kappa^2}{2 \sqrt{\alpha^2 \kappa^4 + 4 \tau^2}}} \,
    {\rm sign} (\tau) , \,
    \frac{\pi}{2}
  \Big) .
\end{eqnarray*}
and the two (equal) minima of $\gamma_{\rm PB} (P_{\rm FS})$ occur for $p$ and $\phi$ given by
\begin{eqnarray*}
  ( p, \phi )
  &=&
  \pm
  \bigg(
    \sqrt{\frac{\sqrt{\alpha^2 \kappa^4 + 4 \tau^2} + \alpha \kappa^2}{2 \sqrt{\alpha^2 \kappa^4 + 4 \tau^2}}} \,
    {\rm sign} (\tau) , \,
    \frac{\pi}{2}
  \bigg) .
\end{eqnarray*}

\subsection{Frenet-Serret Geometric Phase}    \label{subsec:geometric_phase}

Upon substituting Eq.~(\ref{eq:An-Ab-vs-s}) into Eq.~(\ref{eq:d-gamma-geo}) and integrating the resulting expression over $s$, the following expression for $\gamma_{\rm FS} (s)$ is obtained:
\begin{eqnarray}
  \gamma_{\rm FS} (s) &=&
  - \frac{(1 - 2 p^2) \alpha \kappa^2 - 4 p \sqrt{1 - p^2} \, \tau \sin \phi}{\alpha^2 \kappa^4 + 4 \tau^2} \, 2 \tau^2 s
  \nonumber \\ &-&
  \frac{2 p \sqrt{1 - p^2} \, \alpha \kappa^2 \tau \cos \phi}{\alpha^2 \kappa^4 + \tau^2}
  \nonumber \\ && \times
  \Big[
    1 - \cos \Big( \sqrt{\alpha^2 \kappa^4 + 4 \tau^2} \, s \Big)
  \Big]
  \nonumber \\ &+&
  \frac{2 \alpha \kappa^2 \tau}{(\alpha^2 \kappa^4 + 4 \tau^2)^{3/2}} \,
  \sin \Big( \sqrt{\alpha^2 \kappa^4 + 4 \tau^2} \, s \Big)
  \nonumber \\ && \times
  \big[
    (1 - 2 p^2) \tau +
    p \sqrt{1 - p^2} \, \alpha \kappa^2 \sin \phi
  \big] .
  \label{eq:gamma_geo-vs-p-phi}
\end{eqnarray}
It should be noted that $\gamma_{\rm FS} (s)$ consists of a constant term [the term 1 in the expression $1 - \cos( \sqrt{\alpha^2 \kappa^4 + 4 \tau^2} \, s)$, which is introduced in order to satisfy the initial condition $\gamma_{\rm FS} (0) = 0$], a linear with $s$ term, and two oscillating terms.  The later are periodic with the period $P_s / 2$, where $P_s = 4 \pi/\sqrt{\alpha ^2 \kappa ^4 + 4 \tau ^2}$.

\subsection{Adiabatic limit: $\alpha \to 0$}  \label{subsec:Adiabatic_limit}

To understand why $\gamma (s) = 0$ if $\alpha = 0$, we use Eq.~(\ref{eq:d-gamma-dyn}) for $\gamma_{\rm PB} (s)$ and Eq.~(\ref{eq:d-gamma-geo}) for $\gamma_{\rm FS} (s)$,
and express $d \gamma (s) / d s$ in Eq.~(\ref{eq:gamma=dyn+geo})  as
\begin{equation}
  \frac{d \gamma (s)}{d s} =
  \big( A_{n}^{*} (s) , A_{b}^{*} (s) \big) \,
  \left(
    \begin{array}{cc}
      - \frac{\alpha \kappa^2}{2} & 0
      \\
      0 & \frac{\alpha \kappa^2}{2}
    \end{array}
  \right) \,
  \left(
    \begin{array}{c}
      A_n (s) \\ A_b (s)
    \end{array}
  \right) .
\end{equation}
Hence, as $\alpha$ approaches zero, $d \gamma (s) / d s$ approaches zero and therefore $\gamma (s)$ approaches zero for any $s$. Moreover, at $\alpha = 0$, $\gamma_{\rm FS} (s)$ is given by
\begin{eqnarray}
  \left.\gamma_{\rm FS} (s) \right|_{\alpha \to 0} &=& 2 p \sqrt{1 - p^2} \, \tau \sin \phi \, s ,
  \label{eq:gamma_geo-vs-p-phi_alpha_0}
\end{eqnarray}
and
\begin{equation}
\left. \big\{ \gamma_{\rm PB} (s) + \gamma_{\rm FS} (s) \big\} \right|_{\alpha \to 0} = 0.
\end{equation}

\subsection{Comparison of geometric optical phases with Aharonov-Bohm and Aharonov-Casher phases}   \label{subsec:ABphase-ACphase}

In this section we compare the PBP the FSGP and the total geometric phase with the Aharonov-Bohm (AB) phase and the Aharonov-Casher (AC) phase.  The AB effect \cite{AB_59} is a quantum-mechanical phenomenon in which an electrically charged particle with charge $q$ is affected by an electromagnetic potential despite being confined to a region in which both the magnetic field and electric field are zero \cite{Band-Avishai-QuantumMechanics}. The charged particle acquires a phase shift 
\begin{equation}   \label{eq:AB-phase-closed}
  \varphi_{\rm AB} = \frac{q}{\hbar c} \oint_P \boldsymbol{\mathcal{A}} ({\bf r}) \cdot d {\bf r} 
  = \frac{q}{\hbar} \iint_{\Sigma_P} {\bf B} ({\bf r}) \cdot d \boldsymbol\Sigma
  =  \frac{q}{\hbar} \, \Phi ,
\end{equation}
where $P$ is a closed path that is traversed in the counterclockwise direction around the magnetic flux $\Phi$ encircled by the closed path, $\boldsymbol{\mathcal{A}} ({\bf r})$ is the vector potential and ${\bf B} ({\bf r}) = \nabla \times \boldsymbol{\mathcal{A}} ({\bf r})$ is the magnetic field.  The AB phase, $\varphi_{\rm AB}$, does not depend on the closed path $P$ provided the magnetic flux $\Phi$ encircled remains unchanged as the closed path is varied.  Hence, the AB phase is referred as a {\it topological} phase~\cite{Cohen_19}.

The AC effect \cite{AC_84} occurs when a particle with a magnetic moment moves on a closed path so that the magnetic moment is subject to an electric field, and hence the dynamics of the particle is affected due to spin-orbit coupling.  Let us consider a specific case of the AC effect that arises when a spin-1/2 particle with magnetic moment ${\boldsymbol \mu} = g \mu_B {\boldsymbol \sigma}$ moves on a closed path in the $x$-$y$ plane (e.g., a ring of radius $R$) and the ring is threaded by a perpendicular line along the $z$-axis with homogeneous linear charge density $\lambda$ that generates a radial electric field ${\bf E} = \tfrac{\lambda}{2 R} \hat{{\bf r}}$.  Therefore, a particle with velocity ${\bf v}$ moving along the ring is subject to an effective magnetic field ${\bf B} = \tfrac{{\bf E} \times {\bf v}}{c} = B \, \hat{\bf z}$ in its rest frame.  The particle acquires an AC phase $\varphi_{\rm AC}$ while traversing the closed ring $P$:
\begin{eqnarray} \label{eq:ACP}
  \varphi_{\rm AC} (\sigma) &=&
  \frac{g \mu_B \sigma}{2 \hbar c}
  \oint_{P} \big( E_y ({\bf r}) d x - E_x ({\bf r}) d y \big)
  \nonumber \\ &=&
  - 2 \pi \,
  \frac{g \mu_B \lambda \sigma}{\hbar c} .
\end{eqnarray}
Here the quantum number $\sigma$ is the projection of the spin on the $z$-axis. Note that in this configuration the direction of the effective magnetic field is constant and is along the $z$-axis.  In more general configurations, the Zeeman term ${\bf B}\cdot{\bm \sigma}$ can depend on ${\bf r}$ and the direction of the effective magnetic field varies along the path (for example, when the path is {\it not} restricted to the $x$-$y$ plane) and, in contradistinction to Eq.~(\ref{eq:ACP}), the integral in the expression for the AC phase must be path ordered.  
 
References \cite{AC_84, Dulat_12, Cohen_19} claim that the AC phase, $\varphi_{\rm AC}$, does not depend on the details of the closed path $P$, i.e., the path can be arbitrarily deformed, i.e., it is a {\it topological} phase. Unlike the AB phase and the AC phase, the PBP, the FSGP and the total geometric phase depend on the torsion $\tau$ and/or the curvature $\kappa$ of the helix, hence they are called {\it geometric} phases \cite{Cohen_19}.

\subsection{PBP, FSGP and the total geometric phases of the polarization vector}   \label{subsec:dyn_geo_total_phase}

\begin{figure}
\centering
  \includegraphics[width = 0.9 \linewidth,angle=0] {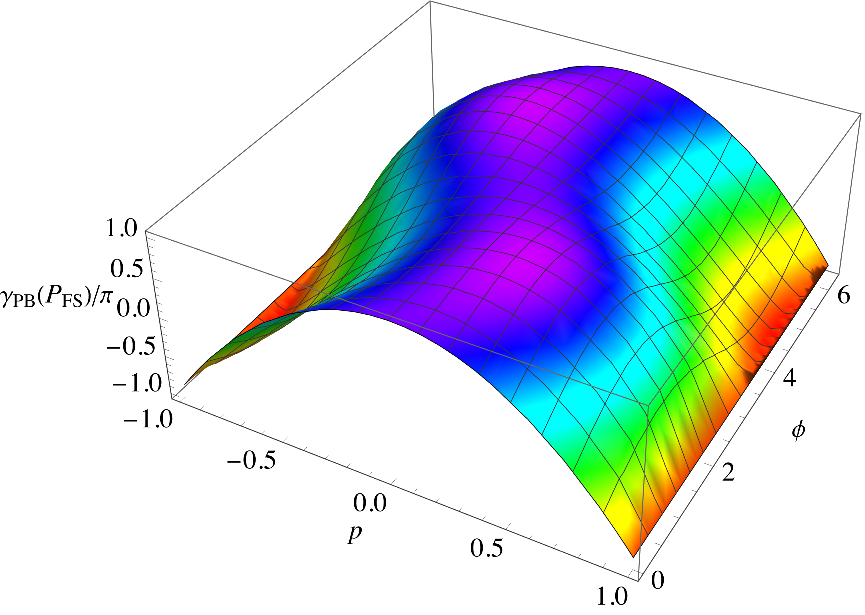}
\caption{\footnotesize  
The PBP $\gamma_{\rm PB}(P_{\rm FS})$ (divided by $\pi$) versus the probability amplitude $p$ and initial angle $\phi$ for parameters $r = 1$, $c = 1/4$ and $\alpha = 1$.
\label{gammadyn_vs_p_phi_r_c}}
\end{figure}

Figure \ref{gammadyn_vs_p_phi_r_c} plots the PBP $\gamma_{\rm PB}(P_{\rm FS})$ versus $p$ and $\phi$ for helix parameters $r = 1$, $c = 1/4$ and $\alpha = 1$. $\gamma_{\rm PB}(P_{\rm FS})$ has a maximum of $3.248$ at $(p,\phi) = (0.2417,\pi/2)$ and at $(p,\phi) = (-0.2417, 3\pi/2)$, and a minimum of $-3.248$ at $(p,\phi) = (0.9703, 3\pi/2)$ and at $(p,\phi) = (-0.9703, \pi/2)$.  Note the symmetry of $\gamma_{\rm PB}(P_{\rm FS})$: transforming $p \to -p$ and $\phi \to -\phi$ leaves $\gamma_{\rm PB}(P_{\rm FS})$ unchanged, as is clear from the figure.

\begin{figure}
\centering
  \includegraphics[width = 0.9 \linewidth,angle=0] {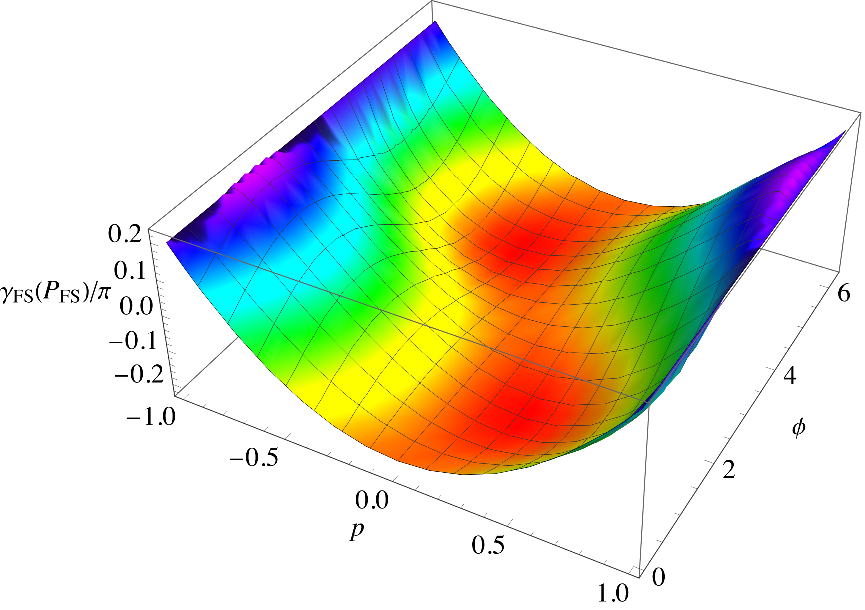}
\caption{\footnotesize  
The FSGP $\gamma_{\rm FS}(P_{\rm FS})$ (divided by $\pi$) versus the probability amplitude $p$ and initial angle $\phi$ for parameters $r = 1$, $c = 1/4$ and $\alpha = 1$.
\label{gammageo_vs_p_phi_r_c}}
\end{figure}

\begin{figure}
\centering
  \includegraphics[width = 0.9 \linewidth,angle=0] {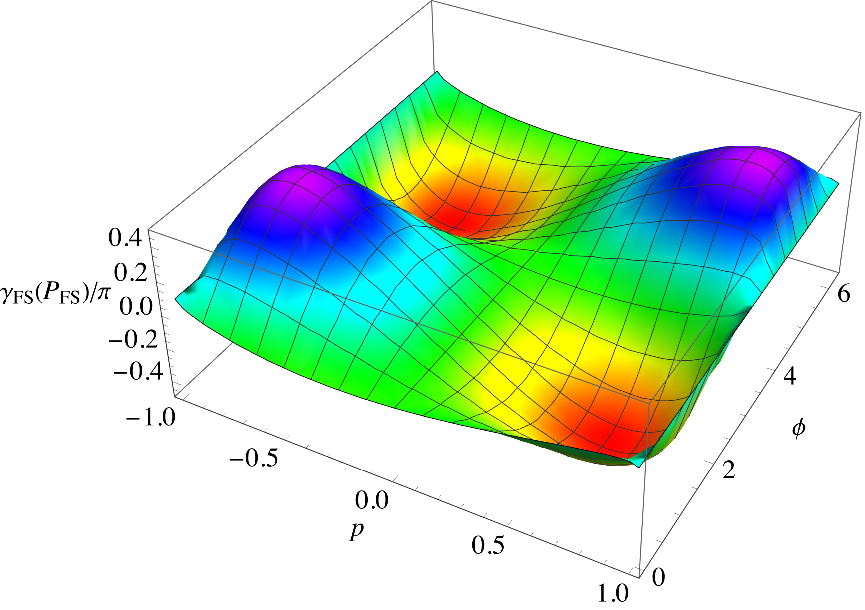}
\caption{\footnotesize  
The FSGP $\gamma_{\rm FS}(P_{\rm FS})$ (divided by $\pi$) versus the probability amplitude $p$ and initial angle $\phi$ for parameters $r = 1$, $c = 1/4$ and $\alpha = 0$.
\label{gammageo_vs_p_phi_r_c_alpha-0}}
\end{figure}

\begin{figure}
\centering
  \includegraphics[width = 0.9 \linewidth,angle=0] {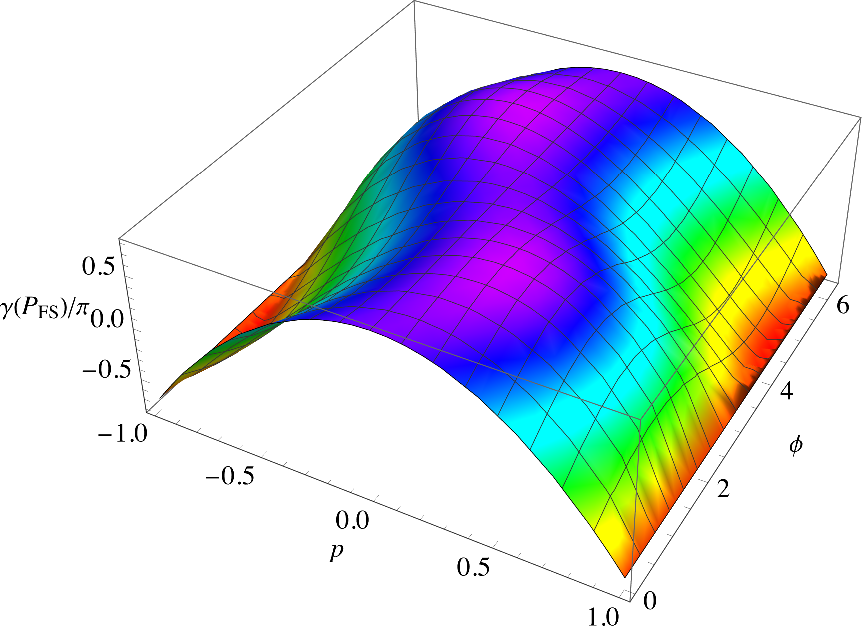}
\caption{\footnotesize  
The total geometric phase $\gamma(P_{\rm FS})$ (divided by $\pi$) versus the probability amplitude $p$ and initial angle $\phi$ for parameters $r = 1$, $c = 1/4$ and $\alpha = 1$.
\label{gamma_vs_p_phi_r_c}}
\end{figure}

Figure \ref{gammageo_vs_p_phi_r_c} plots the FSGP $\gamma_{\rm FS}(P_{\rm FS})$ versus $p$ and $\phi$ for parameters $r = 1$, $c = 1/4$ and $\alpha = 1$. Note that the FSGP is about an order of magnitude smaller than the PBP for $\alpha = 1$.  For comparison, the FSGP $\gamma_{\rm FS}(P_{\rm FS})$ for $r = 1$, $c = 1/4$ and $\alpha = 0$ is plotted in Fig.~\ref{gammageo_vs_p_phi_r_c_alpha-0}.  Clearly the stress-induced birefringence has a significant effect given the differences in the two figures.

Figure \ref{gamma_vs_p_phi_r_c} plots the total geometric phase $\gamma(P_{\rm FS})$ (divided by $\pi$) versus the probability amplitude $p$ and initial angle $\phi$ for parameters $r = 1$, $c = 1/4$ and $\alpha = 1$. The total geometric phase is just the sum of the PBP and the FSGP.  For $\alpha = 1$, the PBP is an order of magnitude larger than the FSGP, hence the total geometric phase looks somewhat like the PBP.

\section{Coherent superposition of the initial and final polarization vectors}   \label{sec:coherent_superposition}

One method of measuring geometric phases is to preform interference experiments (see Ref.~\cite{Malhotra_18} which discusses a non-interferometric approach for probing geometric phases).  Consider a coherent superposition of a fraction of the incident beam at $s = 0$ and the same fraction of the beam at $s = P_{\rm FS}$.  The intensity of the superposition is proportional to
\begin{equation}   \label{eq:Int}
  I_{\rm FS} =  \big| \frac{{\bf A} (0) + {\bf A} (P_{\rm FS})}{2} \big|^{2} =
  \cos^2 \bigg( \frac{\pi \, \sqrt{\alpha^2 \kappa^4 + 4 \tau^2}}{2 \sqrt{\kappa^2 + \tau^2}} \bigg) .
\end{equation}
Comparing the argument of the cosine in the above equation with Eq.~(\ref{eq:dyn_max}) when $s = P_{\rm FS}$, one can see that the argument is equal to $\gamma_{\rm PB,max} (P_{\rm FS})/2$ where
\begin{equation}   \label{eq:dyn_max_PFS}
  \gamma_{\rm PB,max} (P_{\rm FS}) = \frac{\pi \, \sqrt{\alpha^2 \kappa^4 + 4 \tau^2}}{\sqrt{\kappa^2 + \tau^2}} .
\end{equation}
Hence, measuring $ I_{\rm FS}$ can determine $\gamma_{\rm PB,max} (P_{\rm FS})$.

\begin{figure}
\centering
  \includegraphics[width = 0.9 \linewidth,angle=0] {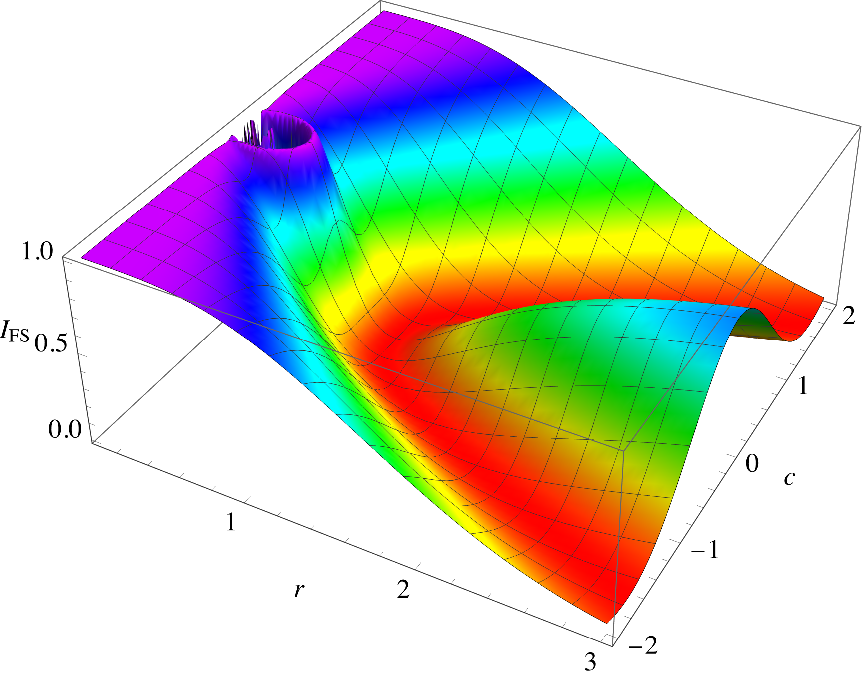}
\caption{\footnotesize  
The intensity $I_{\rm FS}$ plotted as a function of the helix parameters $r$ and $c$ for $\alpha = 1$.  The intensity is independent of initial conditions $p$ and $\phi$.
\label{Intensity_vs_p_ph_r_c}}
\end{figure}

\begin{figure}
\centering
  \includegraphics[width = 0.9 \linewidth,angle=0] {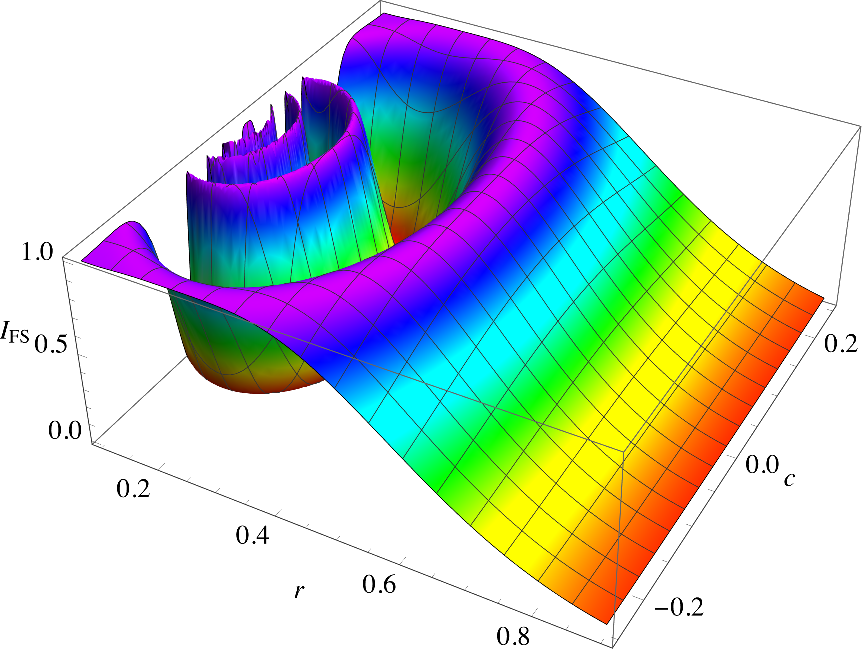}
\caption{\footnotesize  
Blowup of the intensity $I_{\rm FS}$ plotted as a function of the helix parameters $r$ and $c$ for $\alpha = 1$ for small values of $r$ and $c$ showing the fast oscillations of the intensity.
\label{Intensity_vs_p_ph_r_c_blowup}}
\end{figure}

\begin{figure}
\centering
  \includegraphics[width = 0.9 \linewidth,angle=0] {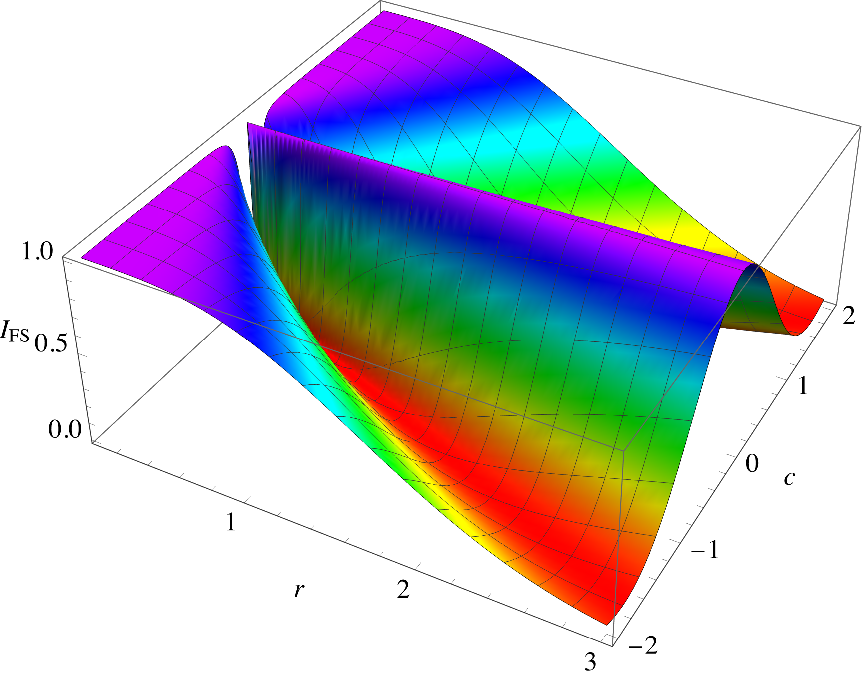}
\caption{\footnotesize  
The intensity $I_{\rm FS}$ plotted as a function of the helix parameters $r$ and $c$ for $\alpha = 0$, see Eq.~(\ref{eq:Int_alpha-0}).
\label{Intensity_vs_p_ph_r_alpha-0}}
\end{figure}

Figure \ref{Intensity_vs_p_ph_r_c} plots $I_{\rm FS}$ defined in Eq.~(\ref{eq:Int}), which ranges from 0 to 1.  As $r$ gets large, $I_{\rm FS} = 0$ on the curve whose asymptote is $|c| = r / \sqrt{3}$, and subsequently approaches $1$ for $c$ far away from the curve. For $r < 1$, $I_{\rm FS}$ goes to 1 for large $|c|$, and oscillates rapidly for small $|c|$.  Figure \ref{Intensity_vs_p_ph_r_c_blowup} shows a blowup of Fig.~\ref{Intensity_vs_p_ph_r_c} in the region of small $r$ and $|c|$ that showcases the rapid oscillations of $I_{\rm FS}$ in this region.  Figure \ref{Intensity_vs_p_ph_r_alpha-0} shows $I_{\rm FS}$ when $\alpha = 0$.  A comparison of Figs.~\ref{Intensity_vs_p_ph_r_c} and \ref{Intensity_vs_p_ph_r_alpha-0} reveals the similarities and differences in the intensity for the case of $\alpha = 0$ and `large' $\alpha$.

The intensity $I_{\rm FS}$ is solely dependent on the parameters $\alpha$, $\kappa$ and $\tau$, and is not dependent on the initial conditions specified by $p$ and $\phi$.  For $\alpha \ne 0$, the phase in Eq.~(\ref{eq:Int}) is not $\gamma_{\rm FS}(P_{\rm FS})$, $\gamma_{\rm PB}(P_{\rm FS})$ or $\gamma(P_{\rm FS})$, but is equal to $\gamma_{\rm PB,max} (P_{\rm FS})/2$.  When $\alpha \to 0$,
\begin{equation}   \label{eq:Int_alpha-0}
  I_{\rm FS} \to  \cos^2 ( \Omega(r, c)/2 ) ,
\end{equation}
where $\Omega(r, c)$ is given by Eq.~(\ref{eq:Omega}).  When $\alpha$ is non-zero, there is a modification of the angle appearing in the expression for the intensity which is due to linear birefringence.

Note that when $\alpha = 0$, $I_{\rm FS}$ is related to both $\gamma_{\rm FS}(P_{\rm FS})$ and $\gamma_{\rm PB}(P_{\rm FS})$ since these phases are directly related to $\Omega$ for $\alpha = 0$.  However, measuring the intensity $I_{\rm FS}$ in an interference experiment will {\it in general} not determine the FSGP, but instead determines the {\it maximum of the PBP}.

\section{Relationship between $\gamma_{\rm FS}$ and $\Omega$}  \label{sec:gammaB-Omega}

Here we compare the solid angle $\Omega (r, c)$ given in
Eq.~(\ref{eq:Omega}) and the PBP $\gamma_{\rm PB} (P_{\rm FS})$
given in Eq.~(\ref{eq:gamma-dyn-Omega-p-phi}) for $s = P_{\rm FS}$.
Note that $\gamma_{\rm PB} (P_{\rm FS})$
depends on the initial conditions parameterized by $p$ and $\phi$, see Eq.~(\ref{eq:initial}), as well as the parameter $\alpha$ (which depends on the fiber radius and elastic properties of the material).  In contradistinction, $\Omega (r, c)$ is a function only of the helix radius $r$ and pitch $2 \pi c$ (it does not depend upon $\alpha$, nor on the initial conditions).

If the three conditions, $(p, \phi) = \pm (1/\sqrt{2}, -\pi/2)$ and  $\alpha = 0$, are satisfied, $\gamma_{\rm FS} (P_{\rm FS}) = -\Omega (r, c)$ and $\gamma_{\rm PB} (P_{\rm FS}) = \Omega (r, c)$.  If the three conditions,  $(p, \phi) = \pm (1/\sqrt{2}, \pi/2)$ and $\alpha = 0$, are satisfied, $\gamma_{\rm FS} (P_{\rm FS}) = \Omega (r, c)$  and $\gamma_{\rm PB} (P_{\rm FS}) = - \Omega (r, c)$.  Only under these conditions are the PBP and the FSGP related to $\Omega(r, c)$.  The condition $\alpha = 0$ implies that (a) no stress-induced birefringence is present, and (b) the light wavenumber $k \to \infty$, i.e., the wavelength $\lambda \to 0$ \cite{Berry_87_N}.

The extrema $\gamma_{\rm dyn, max} (s)$ and $\gamma_{\rm dyn, min} (s)$ for fixed $s$
as functions of $p$ and $\phi$ are given in Eqs.~(\ref{eq:dyn_max}) and (\ref{eq:dyn_min}).  The maximum $\gamma_{\rm dyn, max} (P_{\rm FS})$ calculated for $s = P_{\rm FS}$ is
\begin{equation}
  \gamma_{\rm PB,max} (P_{\rm FS}) =
  \sqrt{1 + \frac{\alpha^2 \kappa^4}{4 \tau^2}} \, \Omega (r, c) .
  \label{eq:dyn_max}
\end{equation}
This equation demonstrates that if $\alpha = 0$, $\gamma_{\rm PB,max} (P_{\rm FS})$
is equal to $\Omega (r, c)$.  However, birefringence that occurs for $\alpha \neq 0$, affects the PBP, and $\gamma_{\rm PB,max} (P_{\rm FS})$ is not in general equal to $\Omega (r, c)$.

\section{Fluctuations of the GPIO due to fluctuations of the helix parameters}  \label{sec:fluctuations}

Experiments to measure the variance of the polarization rotation in optical fibers due to variance in the optical path along the fiber were reported in Ref.~\cite{Ye_90}. In this section we examine the fluctuations of $\gamma_{\rm PB}(\kappa, \tau, \alpha)$ as a result of possible fluctuations in the fiber parameters $r$, $c$ and $\alpha$ for fixed arc-length $s$.  For example, let us assume that there is a Gaussian probability distribution for some of the parameters $r$, $c$ and $\alpha$, with expectation values designated as $r_0$, $c_0$ and $\alpha_0$, and the variances $\sigma_{r}^{2}$, $\sigma_{c}^{2}$ and $\sigma_{\alpha}^{2}$. Assuming that the standard deviations of the variables are smaller than the corresponding mean values, $\sigma_{r} \ll r_0$, $\sigma_{c} \ll c_0$ and $\sigma_{\alpha} \ll \alpha_0$, and neglecting terms of order $\sigma^4$, we find the expectation value $\bar \gamma_{\rm PB}$ of $\gamma_{\rm PB} (\kappa, \tau, \alpha)$ and its variance $\sigma_{\gamma_{\rm PB}}^{2}$:
\begin{eqnarray}
  \bar \gamma_{\rm PB} &=&
  \gamma_{\rm PB} (r_0, c_0, \alpha_0) +
  \frac{\partial^2 \gamma_{\rm PB} (r_0, c_0, \alpha_0)}{\partial r_0^2}
  \sigma_{r}^{2}
  \nonumber \\ &+&
  \frac{\partial^2 \gamma_{\rm PB} (r_0, c_0, \alpha_0)}{\partial c_0^2}
  \sigma_c^2 +
  \frac{\partial^2 \gamma_{\rm PB} (r_0, c_0, \alpha_0)}{\partial \alpha_0^2}
  \sigma_\alpha^2 .
  \quad
  \label{eq:gamma_B-averaged-gen}
\end{eqnarray}
Note that the expectation value $\bar \gamma_{\rm PB}$ depends on the variances of $r$, $c$, and $\alpha$ i.e., $\bar \gamma_{\rm PB} \ne \gamma_{\rm PB} (r_0, c_0, \alpha_0)$.  Hence, average experimental results depend on the variances of $r$, $c$ and $\alpha$.

\section{Summary and Conclusions}  \label{sec:summary}

The concept of geometric phase is fundamental and establishes a deep connection between classical mechanics, quantum mechanics and optics (where it is referred to as the GPIO $\gamma_{\rm FS}$). In this work, we have analyzed the geometric phase $\gamma_{\rm FS}$ for light propagating in a circular helix optical fiber. Using the Frenet-Serret system we obtain the curvature $\kappa$, the torsion $\tau$, and the three orthogonal vectors, the tangent vector ${\bf T}$, the normal vector ${\bf N}$, and the binormal vector ${\bf B}$ along the helical curve. Section~\ref{sec:polarization_helix} presents an analytic expression for the amplitudes $A_n(s)$ and $A_b(s)$ in the polarization vector ${\bf A}(s) = A_n(s) {\bf N}(s) + A_b(s) {\bf B}(s)$ as a function of the arc-length $s$, and the phases $\gamma_{\rm PB}$ and $\gamma_{\rm FS}$ for arbitrary initial conditions of the polarization.  Section~\ref{sec:gammaB-Omega} shows that, even for circularly polarized light, $\gamma_{\rm FS}$ is generally not equal to the solid angle $\Omega(r,c)$ on the sphere subtended by ${\bf T}(s)$ as $s$ moves from $s = 0$ and back to the same point.  For birefringent fibers, the geometric phases depend on initial conditions of the polarization vector and on the parameter $\alpha =\rho^2 a$, which is the product of the square of the radius of the optical fiber, $\rho$ and the constant $a$ which depends on the wavelength of the light and the elastic properties of the fiber.  For fibers with negligible stress-induced birefringence, the geometric phases still depend on the initial conditions for the polarization vector and on the wavelength unless the adiabatic limit, $\alpha \to 0$, is imposed. In contrast, $\Omega(r,c)$ is not dependent on $a$ nor does it depend on the initial conditions for the polarization or the wavelength. The dependence of the mean and standard deviation of the GPIO $\gamma_{\rm FS}$ on fluctuations of the parameters $r, c, \alpha$ is discussed in Sec.~\ref{sec:fluctuations}.  The mean value of $\gamma_{\rm FS}$ is dependent upon the variances of $r$, $c$, and $\alpha$.  Although it is well established that disorder can result in the localization of light, as evidenced by previous research \cite{Wiersma_97, Segev_13}, a comprehensive understanding of the impact of disorder on light polarization and the GPIO remains a topic for further investigation.  It remains to study optical fibers with curvature and torsion that depend on arc-length so that the motion of the polarization vector (or Stokes vector) on the Poincar\'e sphere is not planar because the matrix in Eq.~(\ref{eq:amp_eq}) depends on arc-length.

\section*{Acknowledgements}
We acknowledge useful conversations with Professor Marek Trippenbach and thank Professor Michael V. Berry for useful comments. We dedicate this article to Professor Sergey Geredeskul and his wife Victoria who were murdered in their home in Ofakim, Israel on October 7, 2024 by Hamas terrorists.

\appendix
\section{PBP and FSGP in a helical fiber using eigenvectors}  \label{append:phases}

In this Appendix we propose an alternative way to calculate the PBP in Eq.~(\ref{eq:d-gamma-dyn}) and the geometric phase in Eq.~(\ref{eq:d-gamma-geo}) using the eigenvectors $| \psi_{\pm} \rangle$ of the matrix appearing on the right-hand-side of Eq.~(\ref{eq:amp_eq}), with eigenvalues $\pm k_s$ [see the definition of $k_s$ just below Eq.~(\ref{eq:An-Ab-vs-s})].
The initial polarization vector written in bra-ket notation, $|\psi(0) \rangle = ( A_n(0) , A_0(s) )^{T}$,
can be expressed as a linear combination of $| \psi_{+} \rangle$ and $| \psi_{-} \rangle$:
\begin{equation}
  | \psi (0) \rangle = q_{+} \, | \psi_{+} \rangle + q_{-} \, | \psi_{-} \rangle .
\end{equation}
Here $q_{+}$ and $q_{-}$ are complex probability amplitudes to find the polarization
vector $| \psi (0) \rangle$ in the state $|\psi_{+}\rangle$ and $|\psi_{-}\rangle$,
respectively, and the
polarization eigenvectors are
\begin{subequations}   \label{subeqs:psi_pm}
\begin{eqnarray}
  | \psi_{+} \rangle &=&
  \left(
    \begin{array}{c}
      i \, {\rm sign} (\tau) \, u_{0} \\ \sqrt{1 - u_{0}^{2}}
    \end{array}
  \right) ,
  \label{eq:psi_p}
  \\
  | \psi_{-} \rangle &=&
  \left(
    \begin{array}{c}
      \sqrt{1 - u_{0}^{2}} \\ i \, {\rm sign} (\tau) \, u_0
    \end{array}
  \right) ,
  \label{eq:psi_m}
\end{eqnarray}
\end{subequations}
where ${\rm sign} (\tau)$ is the sign function equal to 1, 0 or $-1$ depending on
whether $\tau$ is positive, zero, or negative, and
\begin{equation}   \label{eq:u0}
  u_0 =
  \sqrt{ \frac{ 2 k_s - \alpha \kappa^2 }{4 k_s}} =
  \sqrt{\frac{\sqrt{\alpha^2 \kappa^4 + 4 \tau^2} - \alpha \kappa^2}{2 \sqrt{\alpha^2 \kappa^4 + 4 \tau^2}}} \, .
\end{equation}
The initial condition for $A_n (0)$ and $A_b (0)$ in Eq.~(\ref{eq:initial})
can be rewritten as
\begin{subequations}   \label{subeqs:ini-eigen}
\begin{eqnarray}
  A_n (0) &=&
  i \, {\rm sign} (\tau) \, u_0 q_{+} +
  \sqrt{1 - u_{0}^{2}} \, q_{-} ,
  \label{eq:ini-An-eigen}
  \\
  A_b (0) &=&
  \sqrt{1 - u_{0}^{2}} \, q_{+} +
  i \, {\rm sign} (\tau) \, u_0 \, q_{-} .
  \label{eq:ini-Ab-eigen}
\end{eqnarray}
\end{subequations}
Note that the first component of $| \psi_{+} \rangle$ is imaginary and the second component of $| \psi_{+} \rangle$ is real.  Therefore, in order to satisfy the initial condition in Eq.~(\ref{eq:initial}),
 both $q_{+}$ and $q_{-}$ must be complex.  After doing some algebra, the following results is obtained:
\begin{eqnarray}
  q_{+} &=&
  - i \, u_0 \, p \,
  {\rm sign} (\tau) +
  \sqrt{1 - u_{0}^{2}} \,
  \sqrt{1 - p^2} \,
  e^{i \phi} ,
  \label{eq:q_plus}
  \\
  q_{-} &=&
  - i \, u_0
  \sqrt{1 - p^2} \,
  e^{i \phi} \,
  {\rm sign} (\tau) +
  \sqrt{1 - u_{0}^{2}} \,
  p .
  \label{eq:q_minus}
\end{eqnarray}
These equations satisfy the condition $|q_{+}|^{2} + |q_{-}|^{2} = 1$.  The solution of Eq.~(\ref{eq:amp_eq}) satisfying the initial conditions in Eq.~(\ref{subeqs:ini-eigen}) can be rewritten as:
\begin{equation}   \label{eq:An-Ab-vs-Ap-Am-s}
 | \psi (s) \rangle =
  q_+ | \psi_+ \rangle \, 
  e^{- i  k_s s} +
  q_- | \psi_- \rangle \, 
  e^{i  k_s s} .
\end{equation}

\subsection{Pancharatnam-Berry Phase}   \label{subsec:dynamic-append}

Substituting Eq.~(\ref{eq:An-Ab-vs-Ap-Am-s}) into Eq.~(\ref{eq:d-gamma-dyn}) and
integrating the resulting expression with respect to $s$ yields
\begin{equation}   \label{eq:gamma-dyn-res}
  \gamma_{\rm PB} (s) =
  \big( |q_{+}|^{2} - |q_{-}|^{2} \big) \, k_s \, s .
\end{equation}
Equation~(\ref{eq:gamma-dyn-res}) indicates that $\gamma_{\rm PB} (s)$ is equal to the probability $|q_{+}|^{2}$ to find the polarization vector in the polarization state $|\psi_{+}\rangle$ times $k_s s$ plus the probability $|q_{-}|^{2}$ to find the polarization vector in the polarization state $|\psi_{-}\rangle$ times $- k_s s$.  Therefore, it is appropriate to refer to $\gamma_{\rm PB} (s)$ as a PBP.  $\gamma_{\rm PB} (s)$ is dependent on the initial conditions parametrized by the complex parameters $q_{+}$ and $q_{-}$ (or $p$ and $\phi$).  The PBP in Eq.~(\ref{eq:gamma-dyn-res}) can be represented as
\begin{equation}   \label{eq:gamma-dyn-Omega}
  \gamma_{\rm PB} (s) =
  \big( |q_{+}|^{2} - |q_{-}|^{2} \big) \,
  \sqrt{1 + \frac{\alpha^2 \kappa^4}{4 \tau^2}} \,
  \tau s .
\end{equation}
Substituting Eqs.~(\ref{eq:period_FS}), (\ref{eq:q_plus}) and (\ref{eq:q_minus}) into Eq.~(\ref{eq:gamma-dyn-Omega}),
Eq.~(\ref{eq:gamma-dyn-Omega-p-phi}) for $\gamma_{\rm PB} (s)$ is obtained.

\subsection{Frenet-Serret Geometric Phase}   \label{subsec:geometric-append}

Substituting Eq.~(\ref{eq:An-Ab-vs-Ap-Am-s}) into Eq.~(\ref{eq:d-gamma-geo}), the following equation is obtained:
\begin{eqnarray*}
  \frac{d \gamma_{\rm FS} (s)}{d s} &=&
  \tau \, \langle \psi (s) | \sigma_y | \psi (s) \rangle
  \nonumber \\ &=&
  2 \, |  \tau  | \, \big( | q_{-} |^{2} - | q_{+} |^{2} \big) \, u_0 \, \sqrt{1 - u_0^2}
  \nonumber \\ &-&
  2 \tau \big( 1 - 2 u_0^2 \big) \, {\rm Im} \big( q_{+}^{*} q_{-} e^{2 i k_s s} \big) .
\end{eqnarray*}
Substituting Eq.~(\ref{eq:u0}) into this equation and integrating the resulting expression with respect to $s$, 
one obtains the following expression for $\gamma_{\rm FS} (s)$:
\begin{eqnarray}
  \gamma_{\rm FS} (s) &=&
  \big(
    | q_{-} |^{2} -
    | q_{+} |^{2}
  \big) \,
  \frac{2 \tau^2 s}{ \sqrt{ 4 \tau^2 + \alpha^2 \kappa^4 } }
  \nonumber \\ &+&
  \frac{2 \tau \alpha \kappa^2}{ 4 \tau^2 + \alpha^2 \kappa^4 } \,
  {\rm Re}
  \big(
    q_{+}^{*} q_{-} e^{2 i k_s s}
  \big) .
  \label{eq:gamma_geo-s}
\end{eqnarray}
If, $s = P_{\rm FS}$ (the arc-length is a Frenet--Serret period) and $\alpha = 0$ (stress-induced birefringence is negligible),
\begin{equation}     \label{eq:gamma_geo-s-simple}
  \gamma_{\rm FS} (P_{\rm FS}) =
  \Big(
    \big| q_{-} \big|^{2} -
    \big| q_{+} \big|^{2}
  \Big) \, \Omega (r, c).
\end{equation}
If, in addition, $|q_{+}| = 1$ then $\gamma_{\rm FS} (P_{\rm FS}) = -\Omega (r, c)$, but if $|q_{-}| = 1$ then $\gamma_{\rm FS} (P_{\rm FS}) = \Omega (r, c)$.

Upon substituting Eqs.~(\ref{eq:Omega}), (\ref{eq:q_plus}) and (\ref{eq:q_minus})
into Eq.~(\ref{eq:gamma_geo-s-simple}), Eq.~(\ref{eq:gamma_geo-vs-p-phi}) for the geometric phase $\gamma_{\rm FS} (s)$ is obtained.

\section{Extrema of $\gamma_{\rm PB} (P_{\rm FS})$}  \label{append-gamma_dyn-max-min}

In this Appendix we find the extrema of $\gamma_{\rm PB} (P_{\rm FS})$ given in Eq.~(\ref{eq:gamma-dyn-Omega-p-phi}) as a function of $p$ and $\phi$. Since finding the extrema of $\gamma_{\rm PB} (P_{\rm FS})$ using Eq.~(\ref{eq:gamma-dyn-Omega-p-phi}) is somewhat complicated (they are nonlinear in the variables $p$ and $\phi$), instead, one can use Eq.~(\ref{eq:gamma-dyn-Omega}) evaluated at $s = P_{\rm FS}$ as a function of $q_{+}$ and $q_{-}$.  Using equations~(\ref{eq:q_plus}) and (\ref{eq:q_minus}), $p$ and $\phi$ can be expressed in terms of $q_{+}$ and $q_{-}$.  Equation~(\ref{eq:gamma-dyn-Omega}) shows that the maximum of $\gamma_{\rm PB} (P_{\rm FS})$ is
\begin{equation}   \label{eq:gamma_dyn-max-append}
  \gamma_{\rm dyn, max} (P_{\rm FS}) =
  \sqrt{1 + \frac{\alpha^2 \kappa^4}{4 \tau^2}} \,
  \Omega (r, c) ,
\end{equation}
and the maximum occurs for $|q_{+}| = 1$ and $|q_{-}| = 0$. The minimum of $\gamma_{\rm PB} (P_{\rm FS})$ is given by 
\begin{equation}   \label{eq:gamma_dyn-min-append}
   \gamma_{\rm dyn, min} (P_{\rm FS}) = - \gamma_{\rm dyn, max} (P_{\rm FS}),
\end{equation}
and occurs for  $|q_{+}| = 0$ and $|q_{-}| = 1$.

The values of $p$ and $\phi$ at which $\gamma_{\rm dyn,} (P_{\rm FS})$ has a maximum
can be found from the equation $| \psi (0) \rangle = q_{+} | \psi_{+} \rangle$.
This implies
\begin{subequations}   \label{subeqs:p-phi-max}
\begin{eqnarray}
  &&i \, q_{+} \, u_0 \, {\rm sign} (\tau) = p ,
  \label{eq1-p-phi-max}
  \\
  &&q_{+} \, \sqrt{1 - u_0^2} = e^{i \phi} \sqrt{1 - p^2} .
  \label{eq2-p-phi-max}
\end{eqnarray}
\end{subequations}
Here $\tau$, $u_0$, $p$ and $\phi$ are real, and $q_{+}$ is complex.
Moreover, $0 < u_0 < 1$, $|p| < 1$, $0 \leq \phi < 2 \pi$ and $|q_{+}| = 1$.
Equation~(\ref{subeqs:p-phi-max}) has two solutions,
\begin{eqnarray*}
  &&
  q_{+} = i ,
  \quad
  p = - u_0 \, {\rm sign} (\tau) ,
  \quad
  \phi = \frac{\pi}{2} ,
  \\
  &&
  q_{+} = - i ,
  \quad
  p = u_0 \, {\rm sign} (\tau) ,
  \quad
  \phi = \frac{3 \pi}{2} .
\end{eqnarray*}

The values of $p$ and $\phi$ at which $\gamma_{\rm dyn,} (P_{\rm FS})$ has a minimum
can be found from the equation $| \psi (0) \rangle = q_{-} | \psi_{-} \rangle$.
This implies
\begin{subequations}   \label{subeqs:p-phi-min}
\begin{eqnarray}
  q_{-} \sqrt{1 - u_0^2} &=& p ,
  \label{eq1-p-phi-min}
  \\
  i q_{-} u_0 {\rm sign} (\tau) &=& e^{i \phi} \sqrt{1 - p^2} .
  \label{eq2-p-phi-min}
\end{eqnarray}
\end{subequations}
This equation has two solutions,
\begin{eqnarray*}
  &&
  q_{-} = {\rm sign} (\tau) ,
  \quad
  p = \sqrt{1 - u_0^2} \, \, {\rm sign} (\tau) ,
  \quad
  \phi = \frac{\pi}{2} ,
  \\
  &&
  q_{-} = - {\rm sign} (\tau) ,
  \quad
  p = - \sqrt{1 - u_0^2} \, \, {\rm sign} (\tau) ,
  \quad
  \phi = \frac{3 \pi}{2} .
\end{eqnarray*}

\end{document}